\documentstyle[12pt]{article}
\begin{document}
\tolerance=5000
\def\pp{{\, \mid \hskip -1.5mm =}}
\def\cL{{\cal L}}
\def\be{\begin{equation}}
\def\ee{\end{equation}}
\def\bea{\begin{eqnarray}}
\def\eea{\end{eqnarray}}
\def\tr{{\rm tr}\, }
\def\nn{\nonumber \\}
\def\e{{\rm e}}
\def\D{{D \hskip -3mm /\,}}

\def\SEH{S_{\rm EH}}
\def\SGH{S_{\rm GH}}
\def\AdS5{{{\rm AdS}_5}}
\def\S4{{{\rm S}_4}}
\def\gfv{{g_{(5)}}}
\def\gfr{{g_{(4)}}}
\def\SC{{S_{\rm C}}}
\def\RH{{R_{\rm H}}}


\  \hfill 
\begin{minipage}{3.5cm}
NDA-FP-73 \\
April 2000 \\
\end{minipage}

\vfill

\begin{center}
{\large\bf BRANE WORLD INFLATION INDUCED BY QUANTUM EFFECTS}

\vfill

{\sc Shin'ichi NOJIRI}\footnote{email: nojiri@cc.nda.ac.jp}, 
{\sc Sergei D. ODINTSOV}$^{\spadesuit}$\footnote{
On leave from Tomsk State Pedagogical University, RUSSIA. \\
\ \hskip 1cm email: odintsov@ifug5.ugto.mx}, \\

\vfill

{\sl Department of Applied Physics \\
National Defence Academy, 
Hashirimizu Yokosuka 239, JAPAN}

\vfill

{\sl $\spadesuit$ 
Instituto de Fisica de la Universidad de 
Guanajuato \\
Apdo.Postal E-143, 37150 Leon, Gto., MEXICO}

\vfill

{\bf ABSTRACT}

\end{center}

We consider brane-world universe where an arbitrary large 
$N$ quantum CFT is 
living on the domain wall. This corresponds to implementing of 
Randall-Sundrum compactification within the context of AdS/CFT 
correspondence. Using anomaly induced effective action for 
domain wall CFT the possibility of self-consistent quantum 
creation of 4d de Sitter wall Universe (inflation) is demonstrated.
In case of maximally SUSY Yang-Mills theory the exact correspondence 
with radius and effective tension found by Hawking-Hertog-Reall is 
obtained. The hyperbolic wall Universe may be induced by quantum effects 
only for exotic matter (higher derivatives conformal scalar) which has 
unusual sign of central charge.

\newpage

The idea \cite{RS} that we live on the brane 
(where four-dimensional gravity is recovered) embedded in 
higher-dimensional spacetime initiated enermous activity in 
the study of brane worlds. 
The attempts to realize the inflationary brane-worlds have 
been made in refs.\cite{CH,K,N,KK,GS,KS,KKM} (and refs.therein). 
However, there is arbitrariness here. It is caused by the fact 
that details of scenario depend 
on the matter content (equation of state), etc. The inflationary 
brane-world scenario realized due to quantum effects of brane 
matter looks more attractive and universal. Such 
idea has been recently suggested in refs.\cite{NOZ,HHR} 
(see also\cite{NOOO}) where addition of conformal
quantum brane matter to complete effective action has been made. 
That corresponds to implementing of RS compactification within 
the context of renormalization group flow in AdS/CFT set-up. 
In particular, using conformal anomaly induced effective action 
the indication to a possibility of quantum creation of de Sitter 
or Anti-de Sitter wall in 5d AdS Universe has been demonstrated 
in ref.\cite{NOZ}. As brane quantum
matter the maximally SUSY Yang-Mills theory on the wall has 
been considered. In ref.\cite{HHR} the role of conformal anomaly 
in inducing of effective tension which is responsible for 
a de Sitter geometry of domain wall has been stressed and 
the corresponding effective tension has been calculated. 
The extensive calculation of (lorentzian) 
graviton correlator has been 
also 
made there. It was shown that domain wall CFT may significally suppress 
the metric perturbations.

In the present Letter we give the details of such proposal 
(addition of brane quantum matter to RS scenario) with 
applications to 4d cosmology in the 
elegant form, using anomaly induced effective action. 
As a result one can consider 
the arbitrary content of CFT living on the wall. Moreover, the 
formalism is applied not only to 4d de Sitter wall but also to 4d
hyperbolic wall in 5d Anti-de Sitter Universe or 4d conformally 
flat Universe. The equivalence of two approaches \cite{NOZ,HHR} 
is explicitly proved: in case of spherical geometry the same 
effective tension and same equation for radius of de Sitter wall 
Universe is obtained. We also show that hyperbolic wall Universe 
may be realized due to quantum effects only for exotic matter 
(higher derivatives scalar). 
 Note that consideration of large $N$ conformal field theory 
living on the domain wall justifies such approach to brane-world 
quantum cosmology as then quantum contribution is essential.
 
We consider the spacetime whose boundary is 4 dimensional 
sphere $\S4$, which can be identified with a D3-brane. 
The bulk part is given by 5 dimensional 
Euclidean anti de Sitter space $\AdS5$ 
\be
\label{AdS5i}
ds^2_\AdS5=dy^2 + \sinh^2 {y \over l}d\Omega^2_4\ .
\ee
Here $d\Omega^2_4$ is given by the metric of $\S4$ 
with unit radius. One also assumes the boundary (brane) 
lies at $y=y_0$ 
and the bulk space is given by gluing two regions 
given by $0\leq y < y_0$. 

We start with the action $S$ which is the sum of 
the Einstein-Hilbert action $\SEH$, the Gibbons-Hawking 
surface term $\SGH$,  the surface counter term $S_1$ 
and the trace anomaly induced action $W$\footnote{For the introduction to
anomaly induced effective action in curved space-time (with torsion), see
section 5.5 in \cite{BOS}.}: 
\bea
\label{Stotal}
S&=&\SEH + \SGH + 2 S_1 + W \\
\label{SEHi}
\SEH&=&{1 \over 16\pi G}\int d^5 x \sqrt{\gfv}\left(R_{(5)} 
+ {12 \over l^2}\right) \\
\label{GHi}
\SGH&=&{1 \over 8\pi G}\int d^4 x \sqrt{\gfr}\nabla_\mu n^\mu \\
\label{S1}
S_1&=& -{3 \over 8\pi G}\int d^4 x \sqrt{\gfr} \\
\label{W}
W&=& b \int d^4x \sqrt{\widetilde g}\widetilde F A \nn
&& + b' \int d^4x\left\{A \left[2{\widetilde\Box}^2 
+\widetilde R_{\mu\nu}\widetilde\nabla_\mu\widetilde\nabla_\nu 
 - {4 \over 3}\widetilde R \widetilde\Box^2 
+ {2 \over 3}(\widetilde\nabla^\mu \widetilde R)\widetilde\nabla_\mu
\right]A \right. \nn
&& \left. + \left(\widetilde G - {2 \over 3}\widetilde\Box \widetilde R
\right)A \right\} \nn
&& -{1 \over 12}\left\{b''+ {2 \over 3}(b + b')\right\}
\int d^4x \left[ \widetilde R - 6\widetilde\Box A 
 - 6 (\widetilde\nabla_\mu A)(\widetilde \nabla^\mu A)
\right]^2 \ .
\eea 
Here the quantities in the  5 dimensional bulk spacetime are 
specified by the suffices $_{(5)}$ and those in the boundary 4 
dimensional spacetime are by $_{(4)}$. 
The factor $2$ in front of $S_1$ in (\ref{Stotal}) is coming from 
that we have two bulk regions which 
are connected with each other by the brane. 
In (\ref{GHi}), $n^\mu$ is 
the unit vector normal to the boundary. In (\ref{W}), one chooses 
the 4 dimensional boundary metric as 
\be
\label{tildeg}
\gfr_{\mu\nu}=\e^{2A}\tilde g_{\mu\nu}
\ee 
and we specify the 
quantities with $\tilde g_{\mu\nu}$ by using $\tilde{\ }$. 
$G$ ($\tilde G$) and $F$ ($\tilde F$) are the Gauss-Bonnet
invariant and the square of the Weyl tensor, which are given as
\footnote{We use the following curvature conventions:
\begin{eqnarray*}
R&=&g^{\mu\nu}R_{\mu\nu} \\
R_{\mu\nu}&=& R^\lambda_{\ \mu\lambda\nu} \\
R^\lambda_{\ \mu\rho\nu}&=&
-\Gamma^\lambda_{\mu\rho,\nu}
+ \Gamma^\lambda_{\mu\nu,\rho}
- \Gamma^\eta_{\mu\rho}\Gamma^\lambda_{\nu\eta}
+ \Gamma^\eta_{\mu\nu}\Gamma^\lambda_{\rho\eta} \\
\Gamma^\eta_{\mu\lambda}&=&{1 \over 2}g^{\eta\nu}\left(
g_{\mu\nu,\lambda} + g_{\lambda\nu,\mu} - g_{\mu\lambda,\nu} 
\right)\ .
\end{eqnarray*}}

\bea
\label{GF}
G&=&R^2 -4 R_{ij}R^{ij}
+ R_{ijkl}R^{ijkl} \nn
F&=&{1 \over 3}R^2 -2 R_{ij}R^{ij}
+ R_{ijkl}R^{ijkl} \ ,
\eea
In the effective action (\ref{W}), with $N$ scalar, $N_{1/2}$ spinor and 
$N_1$ vector fields, $b$, $b'$ and $b''$ are 
\be
\label{bs}
b={(N +6N_{1/2}+12N_1)\over 120(4\pi)^2}\ , \quad 
b'=-{(N+11N_{1/2}+62N_1) \over 360(4\pi)^2}\ , \quad b''=0
\ee
but in principle, $b''$ may be changed by the finite 
renormalization of local counterterm in gravitational 
effective action. As we shall see later, the term proportional 
to $\left\{b''+ {2 \over 3}(b + b')\right\}$ in (\ref{W}), and 
therefore $b''$, does 
not contribute to the equations of motion (see (\ref{slbr1})). 
For ${\cal N}=4$ $SU(N)$ SYM theory $b=-b'={N^2 -1 \over 4(4\pi )^2}$. 
We should also note that $W$ in (\ref{W}) is defined up to 
conformally invariant functional, which cannot be determined 
from only the conformal anomaly. The conformally flat space is a pleasant 
exclusion where anomaly induced effective action is defined uniquely.
However, one can argue that such conformally invariant functional gives 
next to leading contribution as mass parameter of regularization 
may be adjusted to be arbitrary small (or large)..

The metric of $\S4$ with the unit radius is given by
\be
\label{S4metric1}
d\Omega^2_4= d \chi^2 + \sin^2 \chi d\Omega^2_3\ .
\ee
Here $d\Omega^2_3$ is described by the metric of 3 dimensional 
unit sphere. If we change the coordinate $\chi$ to 
$\sigma$ by 
\be
\label{S4chng}
\sin\chi = \pm {1 \over \cosh \sigma} \ , 
\ee
one obtains
\be
\label{S4metric2}
d\Omega^2_4= {1 \over \cosh^2 \sigma}\left(d \sigma^2 
+ d\Omega^2_3\right)\ .
\ee
On the other hand, the metric of the 4 dimensional flat 
Euclidean space is given by
\be
\label{E4metric}
ds_{\rm 4E}^2= d\rho^2 + \rho^2 d\Omega^2_3\ .
\ee
Then by changing the coordinate as 
\be
\label{E4chng}
\rho=\e^\sigma\ , 
\ee
one gets
\be
\label{E4metric2}
ds_{\rm 4E}^2= \e^{2\sigma}\left(d\sigma^2 + d\Omega^2_3\right)\ .
\ee
For the 4 dimensional hyperboloid with the unit radius, 
the metric is given by
\be
\label{H4metric1}
ds_{\rm H4}^2= d \chi^2 + \sinh^2 \chi d\Omega^2_3\ .
\ee
Changing the coordinate $\chi$ to $\sigma$  
\be
\label{H4chng}
\sinh\chi = {1 \over \sinh \sigma} \ , 
\ee
one finds
\be
\label{H4metric2}
ds_{\rm H4}^2 = {1 \over \sinh^2 \sigma}\left(d \sigma^2 
+ d\Omega^2_3\right)\ .
\ee
We now discuss the 4 dimensional hyperboloid whose boundary 
is the 3 dimensional sphere $S_3$ but we can consider the cases that 
the boundary is a 3 dimensional flat Euclidean space $R_3$ or a 
3 dimensional hyperboloid $H_3$. We will, however, only consider 
the case that the boundary is $S_3$ since the results for other 
cases are almost equivalent.

Motivated by (\ref{AdS5i}), (\ref{S4metric2}), 
(\ref{E4metric2}) and (\ref{H4metric2}), one assumes 
the metric of 5 dimensional space time as follows:
\be
\label{metric1}
ds^2=dy^2 + \e^{2A(y,\sigma)}\tilde g_{\mu\nu}dx^\mu dx^\nu\ ,
\quad \tilde g_{\mu\nu}dx^\mu dx^\nu\equiv l^2\left(d \sigma^2 
+ d\Omega^2_3\right)
\ee
and we identify $A$ and $\tilde g$ in (\ref{metric1}) with those in 
(\ref{tildeg}). Then we find $\tilde F=\tilde G=0$, 
$\tilde R={6 \over l^2}$ etc. 
Due to the assumption (\ref{metric1}), the actions in (\ref{SEHi}), 
(\ref{GHi}), (\ref{S1}), and (\ref{W}) have the following forms:
\bea
\label{SEHii}
\SEH&=& {l^4 V_3 \over 16\pi G}\int dy d\sigma \left\{\left( -8 
\partial_y^2 A - 20 (\partial_y A)^2\right)\e^{4A} \right. \nn
&& \left. +\left(-6\partial_\sigma^2 A - 6 (\partial_\sigma A)^2 
+ 6 \right)\e^{2A} + {12 \over l^2} \e^{4A}\right\} \\
\label{GHii}
\SGH&=& {3l^4 V_3 \over 8\pi G}\int d\sigma \e^{4A} 
\partial_y A \\
\label{S1ii}
S_1&=& - {3l^3 V_3 \over 8\pi G}\int d\sigma \e^{4A} \\
\label{Wii}
W&=& V_3 \int d\sigma \left[b'A\left(2\partial_\sigma^4 A
 - 8 \partial_\sigma^2 A \right) \right. \nn
&&\left. - 2(b + b')\left(1 - \partial_\sigma^2 A 
 - (\partial_\sigma A)^2 \right)^2 \right]\ .
\eea
Here $V_3$ is the volume or area of the unit 3 sphere. 

In the bulk, one obtains the following equation of motion 
from $\SEH$ by the variation over $A$:
\be
\label{eq1}
0= \left(-24 \partial_y^2 A - 48 (\partial_y A)^2 
+ {48 \over l^2}
\right)\e^{4A} + {1 \over l^2}\left(-12 \partial_\sigma^2 A 
- 12 (\partial_\sigma A)^2 + 12\right)\e^{2A}\ .
\ee
Then one finds a solution
\be
\label{blksl}
A=\ln\sinh{y \over l} - \ln \cosh\sigma\ ,
\ee
which corresponds to the metric of $\AdS5$ in (\ref{AdS5i}) 
with (\ref{S4metric2}). 
There exists also the solution 
\be
\label{blksl2}
A={y \over l} + \sigma\ ,
\ee
which corresponds to (\ref{E4metric2}), and another solution
\be
\label{blksl3}
A=\ln\cosh{y \over l} - \ln\sinh\sigma\ ,
\ee
corresponds to (\ref{H4metric2}). 
One should note that all the metrics in (\ref{blksl}), 
(\ref{blksl2}) and (\ref{blksl3}) locally describe the 
same spacetime, that is the local region of $\AdS5$, in 
the bulk. As we assume, however, that there is a brane at 
$y=y_0$, the shapes of the branes are different from each 
other due to the choice of the metric. 

On the brane at the boundary, 
one gets the following equation: 
\bea
\label{eq2}
0&=&{48 l^4 \over 16\pi G}\left(\partial_y A - {1 \over l}
\right)\e^{4A}
+b'\left(4\partial_\sigma^4 A - 16 \partial_\sigma^2 A\right) \nn
&& - 4(b+b')\left(\partial_\sigma^4 A + 2 \partial_\sigma^2 A 
 - 6 (\partial_\sigma A)^2\partial_\sigma^2 A \right)\ .
\eea
We should note that the contributions from $\SEH$ and $\SGH$ are 
twice from the naive values since we have two bulk regions which 
are connected with each other by the brane. 
Substituting the bulk solution  (\ref{blksl}) into 
(\ref{eq2}), one obtains
\be
\label{slbr1}
0={48 l^3 \over 16\pi G}\left(\coth {y_0 \over l} 
 - 1\right)\sinh^4 {y_0 \over l} + 24 b'\ .
\ee
Note that eq.(\ref{slbr1}) does not depend on $b$. 
The effective tension of the domain wall is given by 
\be
\label{tF}
\sigma_{\rm eff}={3 \over 4\pi G l}\coth {y_0 \over l}\ .
\ee
As in \cite{HHR},  defining the radius $R$ of the brane in the 
following way
\be
\label{R}
R\equiv l\sinh{y_0 \over l}\ ,
\ee
one can rewrite (\ref{slbr1}) as 
\be
\label{slbr2}
0={1 \over \pi G}\left({1 \over R}\sqrt{1 + {R^2 \over l^2}}
 - {1 \over l}\right)R^4 + 8b'\ .
\ee
This equation generalizes the corresponding result of ref.\cite{HHR} 
for the case when the arbitrary amount of quantum conformal matter
 sits on de Sitter wall. Adopting AdS/CFT correspondence one can argue
that in symmetric phase the quantum brane matter comes due
to maximally SUSY Yang-Mills theory.

As we have $b'\rightarrow -{N^2 \over 4(4\pi )^2}$ 
in case of the large $N$ limit of ${\cal N}=4$ $SU(N)$ SYM theory, 
we find
\be
\label{slbr3}
{R^3 \over l^3}\sqrt{1 + {R^2 \over l^2}}={R^4 \over l^4}
+ {GN^2 \over 8\pi l^3}\ ,
\ee 
which exactly coincides with the result in \cite{HHR}. 
This equation has the unique solution for positive radius which defines 
brane-world de Sitter Universe (inflation) induced by quantum effects.

On the other hand, if we substitute the solution (\ref{blksl2}), 
corresponding to flat Euclidean brane, 
into (\ref{eq2}), we find that (\ref{eq2}) is always 
(independent of $y_0$) satisfied since 
$\partial_y A={1 \over l}$ and $\partial_\sigma^2 A=0$. 

If one substitutes (\ref{blksl3}), which corresponds to the brane with 
the shape of the hyperboloid, then
\be
\label{slbr1b}
0={48 l^3 \over 16\pi G}\left(\tanh {y_0 \over l} 
- 1\right)\cosh^4 {y_0 \over l} + 24 b'\ .
\ee
We should note that eq.(\ref{slbr1b}) does not depend on $b$ 
again. In order that Eq.(\ref{slbr1b}) has a solution, $b'$ 
must be positive, which conflicts with the case of 
${\cal N}=4$ $SU(N)$ SYM theory or usual conformal matter. 
In general, however,  for some exotic theories, like
higher derivative conformal scalar\footnote{Such higher derivative 
conformal scalar naturally appears in infra-red sector of 
quantum gravity\cite{AMO}.}, $b'$ can be positive and one can assume 
for the moment that $b'>0$ here\footnote{For higher 
derivative conformal scalar $b=-8/120 (4\pi)^2$, 
$b'=28/360(4\pi)^2$.}. 
 Defining the radius $\RH$ of the brane in the 
following way
\be
\label{RH}
\RH\equiv l\cosh{y_0 \over l}\ ,
\ee
one can rewrite (\ref{slbr1}) as 
\be
\label{slbr2b}
0={1 \over \pi G}\left(\pm{1 \over \RH}\sqrt{-1 + {\RH^2 \over l^2}}
 - {1 \over l}\right)\RH^4 + 8b'\ .
\ee
Hence, we showed that quantum, conformally invariant matter 
on the wall, leads to the inducing of inflationary 4d de 
Sitter-brane Universe realized within 5d Anti-de Sitter space 
(a la Randall-Sundrum\cite{RS}). Of course, analytical continuation 
of our 4d sphere to Lorentzian signature is supposed what leads 
to ever expanding inflationary brane-world Universe. 
In 4d QFT (no higher dimensions) such idea of anomaly 
induced inflation  has been suggested long ago in refs.\cite{SMM}.
On the same time the inducing of 4d hyperbolic wall in brane Universe is
highly suppressed
and may be realized only for exotic conformal matter. The analysis of the 
role
of domain wall CFT to metric fluctuations may be taken from results of 
ref.\cite{HHR}. 

It is interesting to note that our approach is quite general. In 
particulary, it is not difficult to take into account the quantum gravity 
effects (at least, on the domain wall). That can be done by using the 
corresponding analogs of central charge for various QGs which 
may be taken 
from beta-functions listed in book \cite{BOS}. In other words, there will 
be only QG contributions to coefficients $b,b'$ but no more changes in
subsequent equations\footnote{For example, for Einstein gravity 
$b=611/120(4\pi)^2$ and $b'=-1411/360(4\pi)^2$.}. 
The next question which deserves careful investigation is about the 
(in)stability of such anomaly driven inflation when it evolves to matter 
dominated Universe. This will be discussed elsewhere.

 \ 

\noindent
{\bf Acknowledgements.} 
The work by SDO has been supported in part by CONACyT (CP, ref.990356) 
and in part by RFBR.
SDO would like to thank members of Instituto de Fisica de la UG for kind 
hospitality and Kim Milton for very helpful discussions.

\end{document}